\documentclass[british,aps,reprint]{revtex4}

\usepackage[T1]{fontenc}
\usepackage[latin9]{inputenc}
\usepackage[letterpaper]{geometry}
\usepackage{color}
\usepackage{babel}
\usepackage{amsmath}
\usepackage{amssymb}
\usepackage[unicode=true,pdfusetitle,
 bookmarks=true,bookmarksnumbered=false,bookmarksopen=false,
 breaklinks=false,pdfborder={0 0 0},pdfborderstyle={},backref=false,colorlinks=true]
 {hyperref}
\hypersetup{
 colorlinks=true,  linkcolor=red, urlcolor=blue, citecolor=blue}

\makeatletter
\@ifundefined{textcolor}{}
{%
 \definecolor{BLACK}{gray}{0}
 \definecolor{WHITE}{gray}{1}
 \definecolor{RED}{rgb}{1,0,0}
 \definecolor{GREEN}{rgb}{0,1,0}
 \definecolor{BLUE}{rgb}{0,0,1}
 \definecolor{CYAN}{cmyk}{1,0,0,0}
 \definecolor{MAGENTA}{cmyk}{0,1,0,0}
 \definecolor{YELLOW}{cmyk}{0,0,1,0}
}

\makeatother

\begin{document}
\selectlanguage{british}

\title{Response to \emph{Comment on ``Time delays in molecular photoionization''}:
\\
Extended Discussion \& Technical Notes}

\author{P. Hockett}
\email{paul.hockett@nrc.ca}

\selectlanguage{british}%

\affiliation{National Research Council of Canada, 100 Sussex Drive, Ottawa, K1A
0R6, Canada}

\author{E. Frumker}

\affiliation{Department of Physics, Ben-Gurion University of the Negev, Beer-Sheva
84105, Israel}
\begin{abstract}
In a comment on our article \emph{Time delays in molecular photoionization}
\cite{Hockett2015c}, Baykusheva \& Wörner reproduce canonical scattering
theory, and assert that our results are inconsistent with this well-established
theory \cite{Baykusheva2016}. We absolutely refute this assertion
and the spirit of the comment, although we do agree with Baykusheva
\& Wörner that the textbook theory is correct. In a short response,
\emph{Response to Comment on ``Time delays in molecular photoionization''
}\cite{Hockett2016d}\emph{,} we have already provided a clear rebuttal
of the comment, but gave no technical details. In this fuller response
we extend those brief comments in the spirit of completeness and clarity,
and provide three clear rebuttals to Baykusheva \& Wörner based on
(1) logical fallacy (category error), (2) theoretical details of the
original article, (3) textural content of the original article. In
particular, rebuttal (1) clearly and trivially points to the fact
that there is no issue here whatsoever, with recourse to theoretical
details barely required to demonstrate this, as outlined in the short
version of our response. Our numerical results are correct and reproduce
known physical phenomena, as discussed in the original article hence,
as careful readers will recognise, the formalism used is canonical
scattering theory, and cannot be anything other. In fact, there is
no new fundamental physics here to dispute whatsoever, and nor was
this the raison d'etre of the original article. Additionally, rebuttal
(2) provides the opportunity to discuss, at length, some of these
textbook aspects of photoionization theory, and we hope this discussion
might be of service to new researchers entering this challenging field.
\end{abstract}
\maketitle

\section*{Preamble}

In our article \emph{Time delays in molecular photoionization} \cite{Hockett2015c},
we discussed Wigner time-delays within the context of canonical scattering
theory. More specifically, we discussed the historic background and
physical concepts (sect. 1), then sketched an outline of canonical
theory in a wavepacket notation (sect. 2), with the aim of providing
a physically-intuitive and paedagogical picture for the general reader;
we then proceeded to illustrate the results of numerical methods making
use of computational tools developed by the scattering and photoionization
communities and, finally, to discuss recent experimental methods which
have the potential to measure angle-resolved Wigner delays. Our aim,
therefore, was to present and discuss our methodology and numerical
results in very general terms to a general audience; to discuss the
core scattering physics; and to demonstrate that Wigner delays can
now be computed readily with modest computational resources using
established and user-friendly tools, with only minimal post-processing
of the results required.

In this case, we saw both the potential interest for a general audience
in the methodology and paradigmatic results (particularly in the attosecond
community, where there has been much recent interest in this topic),
but were also aware of the possibility that an unbalanced presentation
may be viewed as too technical \textendash{} hence unsuitable for
a general physics audience \textendash{} or may be perceived to lack
novelty due to the use of existing techniques from scattering theory,
even though these techniques have never before been applied to the
computation of Wigner delays (to the best of our knowledge). In this
light, we sought to strike a balanced presentation between the well-known
concepts, the application of numerical methods within this framework,
and a discussion of the results so obtained. This is, of course, standard
practice in writing a manuscript, in which one must always seek to
balance the presentation of new results and developments with background/prerequisite
fundamentals, in a manner which best serves the target reader. Textbook
results were, therefore, not reproduce in detail, but the focus was
placed on general concepts, and how results for Wigner delays in the
case of molecular photoionization can now be obtained from well-established
numerical methods, and illustrations of such results for paradigmatic
cases. This does, of course, place the onus on the reader to follow-up
with references if they wish to improve their background knowledge.
To balance this somewhat, additional information was provided in supplementary
documents and data files for the interested or expert reader (available
online via Figshare at \url{http://dx.doi.org/10.6084/m9.figshare.2007486}),
which more explicitly detailed the numerical methods used, and their
specification within the usual photoionization theory formalism. The
raw numerical results were also provided directly, for researchers
wishing to work further with this material. The original referees
of the article enjoyed this presentation, and saw the results as useful
for the community.

However, it seems that, perhaps lacking this context, and perhaps
unaware of the supplementary documentation, confusion has arisen,
leading Baykusheva \& Wörner to the conclusion that the standard fundamental
theories underlying our numerical computations were somehow both non-standard
and incorrect. Since we wish to clear-up any such confusion, we present
here three categorical rebuttals to Baykusheva \& Wörner's specific
assertion that ``... the definition given in (1) is not consistent,
neither with the established interpretation of time delay phenomena
{[}2, 3, 4, 5{]} nor the recent theoretical work on photoionization
delays of atomic systems (see, e.g. {[}6{]}).'' \cite{Baykusheva2016}. 

Specifically, our rebuttals are categorised as follows:
\begin{enumerate}
\item \textbf{Logical fallacy (category error):} while we hope that Baykusheva
\& Wörner's technical argument is due purely to an issue of notation
and presentation with our original manuscript, after careful reading
of their comment it seems quite probable to us that the spirit of
the challenge is, in fact, rooted in a category error (\emph{vide
infra}). This seems to be the case since it is trivial to demonstrate
the logical fallacy in their assertion, which appeared to be made
in earnest. Since this argument does not rely on any fundamental knowledge
of scattering theory or photoionization physics, and is a simple application
of logic, we present this first as a robust rebuttal to Baykusheva
\& Wörner's comment, and most readers will find this sufficient to
dispense with the issue. The short ``official'' response to Baykusheva
\& Wörner's comment was based on this analysis \cite{Hockett2016d}.
\item \textbf{Formalism:} technical details of the original article, discussion
of the formalism therein, and extensions providing explicit derivations
for otherwise implicit definitions. In the sense of completeness and
transparency, we include technical details (further discussion and
derivations), as another clear rebuttal of Baykusheva \& Wörner's
technical argument, and in the hope of alleviating any possibility
of notational confusion with the original manuscript.
\item \textbf{Textural content of the original article:} a range of quotes
from the manuscript illustrating statements at odds with Baykusheva
\& Wörner's assertions. Again, we include this for completeness, although
it will not be of interest to most readers.
\end{enumerate}
Since it seems to us that (1) is a rather serious issue, we also take
the opportunity to discuss, at length, some additional aspects of
photoionization theory, including tutorial examples and paedagogy,
and classic papers on the topics, which we hope will serve as additional,
useful material for researchers struggling with the complex, and
somewhat inaccessible, topic of molecular photoionization physics.

\textbf{Note added Jan. 17}: Following this comment, and review by J. Phys. B, Baykusheva \& Wörner issued a slightly updated comment. The original comment remains
on the arXiv (1611.09352v1, \url{https://arxiv.org/abs/1611.09352v1}),
as does our original response (1612.00481v1, \url{https://arxiv.org/abs/1612.00481v1}). Their final comment, and our brief all-author response, will be published in J. Phys. B \cite{Baykusheva2016, Hockett2016d}, the latter including reference to this longer reply for interested readers. We note that the choice of publication of the comment and response was down to the editors and reviewers at J. Phys. B, who decided that the comment and both responses formed ``extremely valuable additions to the literature" and, in particular, ``the second section [technical response] to the rebuttal is necessary, and more helpful to the general reader". However, due to the separate author list on this longer response, and J. Phys. B's 'single response per comment' publication policy, this longer response will not be published directly in J. Phys. B, and becomes (effectively) peer-reviewed online Supplementary Material to the official response, which will be hosted on the arXiv (1612.00481, \url{https://arxiv.org/abs/1612.00481v1}) and Figshare (DOI: 10.6084/m9.figshare.2007486, \url{http://dx.doi.org/10.6084/m9.figshare.2007486}).

\section{Logical Fallacy}

The core scientific assertion of Baykusheva \& Wörner's comment is
that our formalism does not include the dipole matrix element and
full angular dependence of the scattering wave, as defined in canonical
energy-domain scattering theory. They provide a form of the standard
partial wave expansion for photoionization (eqn. 4 in their comment,
further technical details are provided in the following section),
and the Wigner delay $\tau$:

\begin{equation}
f(\epsilon)=\sqrt{\frac{4\pi}{3}}\sum_{l.m}\langle\psi_{lm}|r_{\nu}|\psi_{0}\rangle Y_{lm}(\Omega_{\hat{k}})Y_{1\nu}(\Omega_{\hat{\nu}})\label{eq:fEcanon}
\end{equation}

\begin{equation}
\tau=\frac{d}{d\epsilon}\arg(f(\epsilon))
\end{equation}

Baykusheva \& Wörner conclude their comment with ``Although the spatial
contribution from the phase of the spherical harmonics is discussed
in the original article, the dependence on the matrix elements has
been neglected.'' In other words, following their canonical eqn.
4 (eqn. \ref{eq:fEcanon} herein), they believe our results to be
based on continuum functions of the form:

\begin{equation}
f(\epsilon)=\sqrt{\frac{4\pi}{3}}\sum_{l.m}Y_{lm}(\Omega_{\hat{k}})Y_{1\nu}(\Omega_{\hat{\nu}})\label{eq:fEang}
\end{equation}
or, possibly, 

\begin{equation}
f(\epsilon)=\sqrt{\frac{4\pi}{3}}\sum_{l.m}Y_{lm}(\Omega_{\hat{k}})
\end{equation}
depending on whether one regards the photon angular momentum term
to be part of the omitted dipole matrix element $\langle\psi_{lm}|r_{\nu}|\psi_{0}\rangle$.

Without any detailed knowledge of fundamental scattering or photoionization
physics, one can simply apply logic and physical intuition to see
that this absolutely cannot be the case: without the requisite matrix
elements there would be no energy dependence to our numerical results,
as shown explicitly in eqn. \ref{eq:fEang}, hence the Wigner delay
would show no energy dependence. To be even clearer here: \emph{the
dipole matrix elements define the expansion coefficients for the partial
waves}. If these expansion coefficients are ignored, one is simply
left with an arbitrary sum over a set of states $\{l,m\}$ with unknown
magnitudes and phases, thus selected, presumably, at random by the
researcher, and with a fixed phase relationship defined purely as
a function of angle by the form of the spherical harmonics, i.e. 
\begin{equation}
\arg[f(\epsilon)]\rightarrow\arg[\sum_{l.m}Y_{l,m}(\theta,\phi)]
\end{equation}
This is absolutely contrary to our results: in figures 1 \& 2 clear
angular and energy dependencies are seen in the results, which show
both $\tau_{w}$ and the photoionization cross-sections, and this
complex behaviour was the main focus of the original manuscript. Furthermore,
the Supplementary Materials include full plots, and data, of the energy-dependent
matrix elements used in the calculations.

Yet another form of this statement may be made in the framework of
basic scattering theory: without a radial part to the solution, there
can be no radial asymptote, hence no defined phase to the wavefunction
in the asymptotic (free particle) limit. To restate again more generally:
the well-known phenomena of scattering physics and photoionization
physics (i.e. radial and angular dependence of the outgoing scattering
wave, a defined asymptotic phase, and the energy dependence of the
angular interference pattern of the partial waves etc. etc.), also
observed in our results for the specific case of Wigner delays, \emph{could
not be observed without inclusion of the energy-dependent (radial)
part of the problem}. Throughout our manuscript, we discuss our results
within the expected phenomena of molecular photoionization physics,
including polarization dependence, photoionization cross-sections
(including the features of shape resonances and Cooper minima) and
the underlying scattering physics. None of this physics would be present
in results based on eqn. \ref{eq:fEang}. Sections 3 and 4 of our
manuscript would be entirely redundant in this case. With apologies
to Baykusheva \& Wörner, since we intend no embarrassment, we have
to remark that, on reflection, we are therefore rather surprised by
their assertion, which appears to be a severe, if earnest, category
error rather than simple notational confusion.

In fairness to Baykusheva \& Wörner, one could posit a hypothetical
case where the scattering physics was correctly treated, and continuum
wavefunctions computed, but no dipole integrals taken into account.
In this case, there would be no logical fallacy in their argument:
the calculations would correctly include the scattering physics, but
omit the photoionization physics. Indeed, examples of such continuum
wavefunctions were plotted in fig. 4 of the manuscript to provide
physical insight, and the computation of the scattering solutions
is a prerequisite for the calculation of the dipole matrix elements,
since they are overlap integrals between the initial and final state
wavefunctions. In this hypothetical case, there would be some energy
and angular dependence of the continuum functions, but the photoionization
physics would be missing, hence no coupling of the continuum and bound
state wavefunctions incorporated, and no dipole overlap integrals
between these states would be computed. More precisely, this case
would yield correct continuum states for scattering, but not dipole-prepared
continuum waves corresponding to photoionization, and there would
be no difference to the continuum state as a function of electric-field
polarization in the molecular frame, or ionizing orbital. This case,
however, would not correspond to the to the case Baykusheva \& Wörner
posit, i.e. eqn. \ref{eq:fEang} above, which yields a purely angular
expansion.

\section{Formalism\label{sec:Formalism}}

Let us return to the core scientific assertion of Baykusheva \& Wörner's
comment: that our formalism does not include the dipole matrix element
and full angular dependence of the scattering wave, as defined in
canonical energy-domain scattering theory. As per the above, they
provide a form of the standard partial wave expansion for photoionization
as (eqn. 4 in their comment):

\begin{equation}
f(\epsilon)=\sqrt{\frac{4\pi}{3}}\sum_{l.m}\langle\psi_{lm}|r_{\nu}|\psi_{0}\rangle Y_{lm}(\Omega_{\hat{k}})Y_{1\nu}(\Omega_{\hat{\nu}})\label{eq:fEcanon-2}
\end{equation}

Here, their notation indicates that the scattering wave $f(\epsilon)$
is defined by the matrix element $\langle\psi_{lm}|r_{\nu}|\psi_{0}\rangle$,
and the product of two spherical harmonics $Y_{lm}(\Omega_{\hat{k}})$
and $Y_{1\nu}(\Omega_{\hat{\nu}})$. In this form, the matrix element
is, strictly speaking the generic dipole matrix element, of the form
$\langle\psi_{f}|\mathbf{r.E}|\psi_{i}\rangle$, written specifically
for an initial state $\psi_{0}$, and with both the dipole operator
and the final state expanded in spherical harmonics, to allow separation
of the radial and angular (geometric) parts of the scattering state,
i.e. 
\begin{equation}
\langle\psi_{f}|\mathbf{r}.\mathbf{E}|\psi_{i}\rangle=\langle\psi_{f}|r_{\nu}|\psi_{0}\rangle Y_{1\nu}(\Omega_{\hat{\nu}})=\sum_{lm}\langle\psi_{lm}|r_{\nu}|\psi_{0}\rangle Y_{1\nu}(\Omega_{\hat{\nu}})Y_{lm}(\Omega_{\hat{k}})
\end{equation}

Here the matrix element defines a purely radial integral over the
initial and final state wavefunction, coupled by the raidal part of
the dipole operator - as such this is often termed the \emph{radial
integral} or \emph{radial matrix element} in the literature or, more
generically, the \emph{dipole or photoionization matrix element} (although
this may refer to the full matrix element or the radial matrix element).
If one is to be precise, this expansion assumes a single electron
system (there is no residual ion included in the final state) or,
equivalently, a single active electron. There is also an assumption
of a single polarization geometry, defined by the coordinates $\Omega_{\hat{\nu}}$,
resulting in an absorption of a single photon with one unit of angular
momentum, and projection $\nu$ into the reference frame, implicitly
assumed to be the molecular axis here. To be pedantic, since no angular
integrations are defined, this wavefunction is more completely labelled
as $f(\epsilon,\Omega_{\hat{k}};\,\Omega_{\hat{\nu}})$, fully labelling
the functional dependence on energy and angle, for a specific polarization
geometry. With these stipulations, this equation indeed defines a
valid scattering solution resulting from a dipole interaction. 

It is of note that this type of solution is absolutely standard, has
long been in textbooks \cite{messiah} and in the literature, and
has been written for many different sets of assumptions and with different
notations. Particularly notable and clear derivations, for the specific
cases of atomic and molecular photoionization, have been presented
by Zare and colleagues \cite{Cooper1969,Reid1992,Park1996}, and Dill
\cite{Dill1976}; more recent works have also provided clear derivations
in cases including molecular dynamics \cite{Seideman2001,Stolow2008},
aligned molecular distributions \cite{Underwood2000,Stolow2008} and
for a computationally-tractable simple model \cite{Lucchese2004}.
For clarity, we note that the core photoionization physics here is
identical to the case of photoelectron angular distributions (PADs),
in which the partial wave interferences are manifested in the angle
and energy-dependent photoelectron yields. One of the current authors
has been working in this field for almost a decade and has built significant
expertise in this area; see, for example, refs. \cite{Hockett2007a,Hockett2009,Hockett2011,Hockett2014,Hockett2015,Hockett2015b}.

Throughout our manuscript we are clear on the use of exactly this
standard formalism for our work but, since our aim was not to reproduce
textbook materials, the notation was simplified to an effective partial
wave expansion, vis.:

\begin{equation}
\Psi_{g}=\sum_{lm}\psi_{lm}\label{eq:pwGeneral}
\end{equation}
as given at the beginning of sect. 2. In this case, the notation is
suppressed to a simple set of partial wavefunctions $\psi$, with
$l,m$ labels, to emphasize the core fundamentals of the scattering
state - namely that it is comprised of a set of partial waves, and
different angular momentum states will contribute and interfere (further
discussion of this notational choice is provided below). Later in
the article, in defining the Wigner delays, the same generic partial
wavefunction is more explicitly labled as $\psi_{lm}^{*}(k,\theta,\phi)$.
In the main article we did not, however, explicitly define this wavefunction
with a full partial wave expansion showing the constituent parts of
the photoionization solution, although the contributing factors were
discussed, including scattering phases, dipole integrals and so forth
(again, see below for further discussion), as well as references to
various articles on photoionization and scattering theory. 

However, recognising that the interested or expert reader may wish
to confirm the formalism used, or their general understanding of the
physics, we did include further discussion in the Supplementary Material
(included here as an appendix for completeness), in particular the
following equations:

\begin{equation}
\Psi(\mathbf{r})=\sum_{lm}A_{l}\chi_{l}(r)Y_{lm}(\theta,\phi)=\sum_{lm}\psi_{lm}(\mathbf{r})
\end{equation}
which defines a generic scattering wave expansion, with partial waves
given as $\psi_{lm}(\mathbf{r})$; and for the specific case of photoionization:

\begin{equation}
d(E)=\langle\Phi_{f}(\mathbf{r});\:\Psi(\mathbf{r})|\hat{\mathbf{\mu}}.\boldsymbol{\mathbf{E}}|\Phi_{i}(\mathbf{r})\rangle
\end{equation}

\begin{equation}
\Psi(E,\theta,\phi)=\sum_{lm}d_{l,m}(E)Y_{lm}(\theta,\phi)
\end{equation}
which defines a general dipole matrix element $d(E)$, and the photoionization
wavefunction $\Psi(E,\theta,\phi)$ as an expansion in spherical harmonics,
with coefficients defined by the dipole matrix elements $d_{l,m}(E)$.
We did not provide a full derivation here, but plugging $\Psi(\mathbf{r})$
into the form of $d(E)$, and solving analytically for a specific
polarization geometry (with the dipole operator expanded as angular
momentum coupling terms - see, for example, \cite{Cooper1969,Dill1976}),
will provide a full photoionization solution. In this case, one can
still write a generic partial wave expansion, as per \ref{eq:pwGeneral},
e.g.

\begin{equation}
\Psi(E,\theta,\phi)=\sum_{lm}d_{l,m}(E)Y_{lm}(\theta,\phi)=\sum_{lm}\psi_{lm}(E,\theta,\phi)
\end{equation}
where the partial wavefunction is explicitly a function of energy
and angle, but the dependence on the dipole integrals is implicit.
Again, this is perfectly standard notation.

Finally, and more specifically, we included the formalism provided
by the authors of ePolyScat, which defines the numerical computations
reported in the manuscript (ref. \cite{Toffoli2007}):

\begin{equation}
I_{l,m,\mu}^{p_{i}\mu_{i},p_{f}\mu_{f}}(E)=\langle\Psi_{i}^{p_{i},\mu_{i}}|\hat{d_{\mu}}|\Psi_{f}^{p_{f},\mu_{f}}\varphi_{klm}^{(-)}\rangle
\end{equation}

\begin{equation}
T_{\mu_{0}}^{p_{i}\mu_{i},p_{f}\mu_{f}}(\theta_{\hat{k}},\phi_{\hat{k}},\theta_{\hat{n}},\phi_{\hat{n}})=\sum_{l,m,\mu}I_{l,m,\mu}^{p_{i}\mu_{i},p_{f}\mu_{f}}(E)Y_{lm}^{*}(\theta_{\hat{k}},\phi_{\hat{k}})D_{\mu,-\mu_{0}}^{1}(R_{\hat{n}})\label{eq:ePS-full}
\end{equation}
Full details of the notation are given in the original Supplementary
Material, but suffice to note here that, structurally, this formalism
is again - fundamentally - canonical scattering theory, with a partial
wave expansion applied to the case of photoionization. In the ePolyScat
case, the equations are further defined by point group symmetries
(with the matrix elements symmetrized over sets of spherical harmonics
as appropriate), and include a Wigner rotation matrix element for
the polarization geometry, thus allowing any set of Euler angles to
define this physical parameter. The radial matrix elements denoted
$I_{l,m,\mu}^{p_{i}\mu_{i},p_{f}\mu_{f}}(E)$ are output by the ePolyScat
calculations, and post-processed according to eqn. \ref{eq:ePS-full},
the Wigner delay is then calculated in the standard manner (see below).

To belabour the point - all of these formalisms are effectively identical,
as far as the fundamental physics is concerned, and within a given
set of approximations. There is no new physics here, and fundamental
results of this type have been well-established since the 1950s in
general, and since the 1960s and early 1970s for some of the specific
photoionization cases. Baykusheva \& Wörner comment that ``We argue
that the denition given in (1) {[}our original manuscript{]} is not
consistent, neither with the established interpretation of time delay
phenomena {[}2, 3, 4, 5{]} nor the recent theoretical work on photoionization
delays of atomic systems (see, e.g. {[}6{]}).'' In fact, notational
differences aside, these various forms of the photoionziation matrix
element, the resultant scattered wave, and the total (asymptotic)
phase, are essentially identical. Therefore, the phase content of
the final wavefunction is the same, and will provide the same results
for the Wigner delay. These results are absolutely canonical, and
absolutely consistent with both the established interpretations of
time delay phenomena and the recent work cited by Baykusheva \& Wörner.
For a more thorough touchstone review, which includes time delays
for a range of types of scattering problem, including wavepackets,
we refer the reader to ref. \cite{DeCarvalho2002}. For a more thorough
review of scattering theory, including various distinct energy and
time-domain (wavepacket) frameworks, and the equivalence of the phase
information of same, see ref. \cite{RTscattering}.

In practice, it was effectively the final flavour of these formalisms
that was applied in the numerical calculations, which made use of
ePolyScat results, as described in sect. 3.A. of the original manuscript
(emphasis added):
\begin{quote}
``Continuum wavefunctions and \emph{dipole matrix elements} were
computed with ePolyScat, for the highest-lying $\sigma$-orbitals
in both cases, for linearly polarized ionizing radiation in both parallel
and perpendicular geometries, and for photoelectron energies from
1 to 45~eV. The \emph{phase information from the raw matrix elements},
expressed in terms of angular momentum channels, provides the full
scattering phase-shift, and application of eqn. 3 provides $\tau_{w}$
for each channel. Similarly, eqn. 4 provides the group, or photoelectron
wavepacket, delay. In the calculations, \emph{radial integrals} are
evaluated for $r_{max}=$~10~Å, defining an effective range to the
interaction at which the total phase (hence delay) is defined. By
calculating the \emph{photoionization matrix elements} for a range
of photoelectron energies, the energy dependence of the process can
be mapped out, and the complete dependence of the Wigner delay $\tau_{w}(k,\theta,\phi)$
obtained.''
\end{quote}
Again, although no formalism is given explicitly here, we believe
that most readers will follow the clear content and logic of the text.

To be yet more explicit, one can write the Wigner delay for this formalism,
starting from the final simplified form given in the original manuscript
and substituting in the full expression for the final state, as follows:

\begin{eqnarray}
\tau_{w}^{g}(k,\theta,\phi) & = & \hbar\frac{d\arg(\sum_{l,m}\psi_{lm}^{*}(k,\theta,\phi))}{d\epsilon}\\
\rightarrow\tau_{w,\,\mu_{0}}^{g,p_{i}\mu_{i},p_{f}\mu_{f}}(\epsilon,\theta_{\hat{k}},\phi_{\hat{k}};\,R_{\hat{n}}) & = & \hbar\frac{d}{d\epsilon}\arg\left(\sum_{l,m,\mu}I_{l,m,\mu}^{p_{i}\mu_{i},p_{f}\mu_{f}}(\epsilon)Y_{lm}^{*}(\theta_{\hat{k}},\phi_{\hat{k}})D_{\mu,-\mu_{0}}^{1}(R_{\hat{n}})\right)
\end{eqnarray}
In this form, the unsummed indicies from the ePolyScat results (which
pertain to the specific ionization channel(s) present) remain explicit,
the results are further labelled to correspond to a specific polarization
geometry, as defined by the set of Euler angles $R_{\hat{n}}$, and
incident photon projection in the lab. frame ($\mu_{0}$). For clarity,
the energy dependence has been labelled with $\epsilon$, instead
of denoted by the photoelectron momentum $k$. Additional normalization
factors are included within $I_{l,m,\mu}^{p_{i}\mu_{i},p_{f}\mu_{f}}(\epsilon)$.
Although this form appears complicated, we note that we have significant
experience with the use of ePolyScat and post-processing of the results,
and recently released the first version of a software suite, ePSproc,
for such purposes. This software evolved from codes developed over
a number of years, and carefully tested both in multiple applications.
For the ePSproc release, the code was further tested against benchmark
results provided by the authors of ePolyScat, for the computation
of molecular frame photoelectron angular distributions - an almost
identical problem to that discussed herein; further details, including
the software, examples and documentation, may be found in refs. \cite{Hockett2016a,Hockett2016c}.

The Wigner delays given here can be compared with eqns. (3) \& (5)
of Baykusheva \& Wörner - again, canonical results - given in their
notation as:

\begin{eqnarray}
\tau & = & \frac{d}{d\epsilon}\arg(f(\epsilon))\\
\tau(k,\theta,\phi,\Omega_{\hat{\nu}}) & = & \hbar\frac{d}{d\epsilon}\arg\left(\sqrt{\frac{4\pi}{3}}\sum_{l.m}\langle\psi_{lm}|r_{\nu}|\psi_{0}\rangle Y_{lm}(\Omega_{\hat{k}})Y_{1\nu}(\Omega_{\hat{\nu}})\right)
\end{eqnarray}
where $f(\epsilon)$, as defined in eqn. \ref{eq:fEcanon-2}, has
been explicitly expanded in the full form. Again, this is seen to
be functionally identical (apart from some assumptions and notational
differences) to our formalism. It is, again, an absolutely standard
result.

\textbf{Note added Jan. 2017}: Following this response, Baykusheva
\& Wörner provided a slightly updated comment, with additional eqns.
6 - 8. Specifically, they provide forms of eqn. 15 and 18 herein ,
effectively asserting again, but more categorically, that: 

\begin{equation}
f(\epsilon)\neq\Psi_{g}\label{eq:fE_psiG}
\end{equation}
where $f(\epsilon)$ is the general scattering function in their notation,
while $\Psi_{g}$ is the generic continuum wavefunction in our notation.
This is a more specific version of their original argument, but the
core thrust is unchanged and again, as detailed above, we stress that
these quantities are essentially identical. This notational misunderstanding
is further obfuscated by their new discussion around their eqn. 8:
\begin{quote}
``In the special case of a non-degenerate, real-valued initial-state
function, one finds
\end{quote}
\[
\arg(\langle\psi_{lm}(k)|r|\psi_{0}\rangle)=\arg(\psi_{lm}^{*}(r,k))\,\,\,(8)
\]

\begin{quote}
Therefore, the delays defined in Eqs. (6) and (7) only become equivalent
in situations where a single partial wave (single value of $l$) contributes
to the photoionization process, which is practically never the case
in molecular photoionization. In general, the definitions given in
eq. (6) and (7) involve taking the argument of a sum of complex terms,
which, albeit having equal phases, in general possess different amplitudes.
Thus, the final results obtained from eq. (6) and (7) will not be
equivalent in general.''
\end{quote}
Again, this, effectively, just another statement of eqn. \ref{eq:fE_psiG},
apparently resulting from basic notational misunderstanding. However,
rather than addressing the original fallacy of the comment, this additional
material merely serves to create further knots and confusion, since
it posits some special case where the formalisms are identical, thus
implying they are otherwise distinct, and the use of an apparently
technical argument in this manner serves to provide the patina of
physical legitimacy to the comment, despite a complete lack of physical
foundation. Finally, Baykusheva \& Wörner state:
\begin{quote}
``The discrepancy between Eqs. (1) and (5) can however be resolved
by replacing the definition of $\psi_{lm}$ as continuum partial-wave
\emph{functions} with the definition of $\psi_{lm}$ as partial-wave
\emph{matrix elements}. This is apparently what the authors of Ref.
{[}1{]} have done in their numerical illustrations of Eq. (1).''
\end{quote}
This is again a statement founded on profound misunderstanding, namely
that the dipole-prepared continuum wavefunction does not result from
photoionization. The dipole-prepared continuum wavefunction absolutely
contains the dipole matrix elements (as defined in, e.g. eqns. 9 -
11 herein ), as it must since these provide the amplitudes and phases
for the dipole coupling of the initial state and the continuum state(s);
see discussion in Sect. 1. (However, Baykusheva \& Wörner are correct
on one point - this is indeed what was done in our numerical calculations,
since it is the correct approach.)

Indeed, it seems to us, that the driving force here is simply to publish
and claim originality. Again, we can clarify - as detailed above -
that all of these notations are equivalent, and there is no issue
here whatsoever, or at least nothing beyond basic physical misunderstandings.
Again, we refer the interested reader who wishes to delve into the
full textbook level of detail on this topic, to ref. \cite{RTscattering}
for a thorough and careful discussion of scattering physics in both
stationary state (energy domain) and wavepacket (time domain) forms.
Again, we remark that there is no new physics here, only new misunderstandings,
wilful or otherwise.

\section{Textural specifics}

Although the above discussions are more than sufficient to dispense
with Baykusheva \& Wörner's contention, in the spirit of completeness
and transparent discussion, we note here some additional points from
our manuscript which also attest to the aims and content of the work.
As noted above, throughout our manuscript we are clear on the use
of a standard photoionization formalism for our work, and discussed
this in a time-domain (wavepacket) picture, which is formally equivalent
to the more usual energy-domain formalism. In particular:
\begin{itemize}
\item As stated in the introduction, in reference to scattering phases:
``These results are most simply derived in a stationary state (energy-domain)
picture of scattering, but a wavepacket (time-domain) treatment yields
the same essential features \cite{RTscattering}. Hence, in a time-domain
picture of photoionization, the scattering phase-shift and associated
time delay can be viewed as a group delay of the outgoing photoelectron
wavepacket, born at a time t0 within the ionizing laser pulse.''
\item Also from the introduction: ``More traditional scattering theory
approaches are usually time-independent and most suited to the weak
field regime, hence are appropriate for the consideration of the intrinsic
Wigner delay of the scattering system. Such approaches often use a
\emph{partial-wave} formalism, which allows separation into \textquotedbl{}geometric\textquotedbl{}
and \textquotedbl{}dynamical\textquotedbl{} parts. In this case much
progress can be made analytically, and a deep physical insight into
the characteristics of the scattering can be gained (see, for example,
ref. \cite{Dill1976}). However, to obtain a complete solution to
a complex scattering problem numerical methods are still ultimately
needed for the dynamical part, and a specific formalism for the scattering
system of interest is usually constructed in order to yield tractable
equations (see, for example, refs. \cite{Ivanov2012,Serov2013});
solving molecular scattering problems is therefore non-trivial for
even the simplest cases. This problem can, however, be addressed via
the use of variational techniques to solve the numerical part of the
problem \cite{Lucchese1986}, allowing for a methodology which retains
the full physical insights of scattering theory and the generality
of fully-numerical approaches, but at a significantly lower computational
cost.\\
In this work, we investigate Wigner delays from molecular ionization
based on this general approach.''
\item The use of a time-domain formalism allows for a physically intuitive
picture of the problem (see discussion in ref. \citep{Tannor2007}
for example) and, as detailed in the canonical textbook by Rodberg
and Taylor (ref. \cite{RTscattering} as above), is formally identical.
Rodberg and Taylor discuss this at length, and present numerous energy
and time-domain derivations of canonical scattering theory, and the
interested reader is referred to that seminal text for details.
\item Many other authors have used a similar formalism, as pointed out at
the start of sect. 2: ``As discussed by Wigner \cite{Wigner1955},
Smith \cite{Smith1960} and, more recently, in some depth by various
authors \cite{DeCarvalho2002,Dahlstrom2012,Pazourek2014}, the phase
of the scattering wavefunction can be associated with a time delay
of the outgoing wavepacket, $\Psi_{g}$. In a partial-wave decomposition,
$\Psi_{g}$ is expressed as a coherent sum over partial-waves, $\Psi_{g}=\sum_{lm}\psi_{lm}$.
Here each component is defined by the quantum numbers $(l,m)$, the
electronic orbital angular momentum and its projection onto a given
quantization axis respectively, and each $(l,m)$ pair defines a partial-wave
scattering channel.''\\
``Here $\eta_{g}$ represents the total (group) scattering phase,
determined from $\Psi_{g}$, hence from the coherent summation over
the partial-wave channels.''
\item The dipole matrix elements (radial integrals) can be determined numerically
from ePolyScat, as discussed in sect. 3.1: ``The phase information
from the raw matrix elements, expressed in terms of angular momentum
channels, provides the full scattering phase-shift, and application
of eqn. 3 provides $\tau_{w}$ for each channel. Similarly, eqn. 4
provides the group, or photoelectron wavepacket, delay. In the calculations,
radial integrals are evaluated for $r_{max}=$~10~Å, defining an
effective range to the interaction at which the total phase (hence
delay) is defined. By calculating the photoionization matrix elements
for a range of photoelectron energies, the energy dependence of the
process can be mapped out, and the complete dependence of the Wigner
delay $\tau_{w}(k,\theta,\phi)$ obtained.\\
In the following, we present and discuss these results for the general
reader. Supplementary materials, including additional technical details
of the results, e.g. channel-resolved dipole matrix elements, which
may be of interest to some readers, are available online via Figshare
at \url{http://dx.doi.org/10.6084/m9.figshare.2007486}.''
\item The formalism for the ePolyScat results and photoionization matrix
elements referred to above was given more explicitly in the Supp.
Mat., wherein the details of the scattering calculations and results
obtained from ePolyScat are discussed. The details of this discussion
have already been sketched out above, and are also reproduced below
for completeness. In this case, the dipole matrix element - equivalent
to the term $\langle\psi_{lm}|r|\psi_{0}\rangle Y_{1\nu}(\Omega_{\nu})$
given by Baykusheva \& Wörner - is defined as $d(E)$, and the full
asymptotic wavefunction is then given by $\Psi(E,\theta,\phi)=\sum_{lm}d_{l,m}(E)Y_{lm}(\theta,\phi)$.
Note that the matrix element includes integration over the radial
coordinate $r$. The equivalent equation for the numerical results,
as defined by the authors of ePolyScat \cite{Toffoli2007}, is given
as $T_{\mu_{0}}^{p_{i}\mu_{i},p_{f}\mu_{f}}(\theta_{\hat{k}},\phi_{\hat{k}},\theta_{\hat{n}},\phi_{\hat{n}})=\sum_{l,m,\mu}I_{l,m,\mu}^{p_{i}\mu_{i},p_{f}\mu_{f}}(E)Y_{lm}^{*}(\theta_{\hat{k}},\phi_{\hat{k}})D_{\mu,-\mu_{0}}^{1}(R_{\hat{n}})$,
where $I_{l,m,\mu}^{p_{i}\mu_{i},p_{f}\mu_{f}}(E)$ are, again, the
radial dipole integrals defined as $I_{l,m,\mu}^{p_{i}\mu_{i},p_{f}\mu_{f}}(E)=\langle\Psi_{i}^{p_{i},\mu_{i}}|\hat{d_{\mu}}|\Psi_{f}^{p_{f},\mu_{f}}\varphi_{klm}^{(-)}\rangle$.
\item Regardless of an energy or time-domain picture, the asymptotic phase
of the photoelectron wavefunction is given by the scattering phase.
In other words, $\arg(\Psi_{g})$ is given by the scattering phase
$\eta_{g}$, which is determined by the scattering of the electron
from the molecule (see ref. \cite{Park1996} for detailed discussion).
In the case of photoionization - a ``half-collision'' - in the weak
field limit, the composition of the outgoing scattering wave is described
by a dipole matrix element, which couples the ionizing orbital to
the continuum wavefunction. This matrix element defines the scattering
phases. As discussed above, without this matrix element there would
be no energy dependence to the scattering state, nor any dependence
on the ionization dynamics, since the continuum wave would not be
coupled to an ionizing orbital - ionization physics could not be modelled.
This physical concept, and consequent details, are stressed multiple
times throughout the manuscript, vis. {[}emphasis added{]}:
\begin{itemize}
\item ``In atomic ionization, the relatively simple nature of the scattering
potential results in a continuum wavepacket with little spatial structure,
which can often be described by just two partial-wave channels. In
molecular ionization, the anisotropic nature of the potential means
that many more partial-waves are required to describe the photoelectron
wavepacket, and significant spatial and energy structure is expected.
In essence, the angular structure of the photoelectron wavepacket
is the result of the angular interferences between the partial-waves
at a given energy, while the difference in the dependence of the\emph{
phase-shift of any given $l$-wave on the photoelectron kinetic energy
results} in the strong energy-dependence of the photoionization cross-section
and $\tau_{w}$.'' (Sect. 2) 
\item `` In these plots the surface topography follows the magnitude of
the dipole matrix element (proportional to the square-root of the
photoionization cross-section), while the colour-map shows the energy
and angle-resolved Wigner time.'' (Sect. 3.B) 
\item ``In both cases, the ionizing orbital is the valence $\sigma$-bonding
orbital, with lobes oriented along the molecular axis. The choice
of polarization of\emph{ the ionizing radiation} - either parallel
or perpendicular to the molecular axis - defines the symmetry of the
ionization continuum accessed, hence the symmetry of the continuum
photoelectron wavefunction.'' (Sect. 3.B) 
\item ``Physically, the peaks in the cross-section correspond\emph{ to
maxima in the dipole integrals which define the coupling between initial
orbital and final continuum wavefunctions induced by ionizing radiation,
with an angular dependence given by the partial-wave interferences}.
For $N_{2}(\Sigma_{u})$ this peak is the well-known shape-resonance
\cite{Lucchese1986,Shigemasa1995,Hikosaka2000}, corresponding to
an enhancement of the $l=3$ partial-wave, which can be considered
as a trapping of this part of the outgoing wavepacket due to the form
of the molecular potential energy surface.'' (Sect. 4, telling entitled
``Scattering Dynamics'') 
\item ``In the case of $CO$ the picture is quite different. Here the Wigner
delays are predominantly negative, indicating a slight net repulsive
effect from the molecular potential, and the results are highly asymmetric,
consistent with the loss of inversion symmetry and the form of the
ionizing orbital for a polar diatomic. The repulsive nature of the
potential is most significant at the oxygen end of the molecule, where
the extent of the ionizing orbital is much reduced relative to the
carbon end.'' (Sect. 4) 
\item ``This can be understood by consideration of the radial part of the
continuum wavefunction: at higher energies the photoelectron wavelength
becomes shorter, and the continuum function will become more penetrating
relative to the core wavefunction. Consequently, \emph{the spatial
overlap integral }will incorporate more bound-state density closer
to the core, which is effectively more strongly bound...'' (Sect.
4) 
\item ``Visualization of the scattering wavefunctions provides additional
physical insight into the dynamics of the process. Figure \ref{fig:Continuum-wavefunctions}
shows a selection of continuum wavefunctions at different photoelectron
energies, chosen to represent the evolution of the scattering wavefunctions
towards the peak in the cross-sections (shape-resonance), with symmetries
concomitant with ionization parallel to the molecular frame ($N_{2}(\Sigma_{u})$
and $CO(\Sigma)$).'' (Sect. 4) 
\item ``with the observed continuum structure corresponding to the rise
and fall of the $l=3$ partial-wave component over this energy range,
including a significant change in the magnitude of the wavefunction
in the core region which has a strong effect on the overall ionization
yield'' (Sect. 4) 
\item ``In all cases, the asymptotic phase-shift of the waves is approximately
established at the length-scales shown ($r_{max}=$10~Å), and phase
differences can be observed in the plots.'' (Sect. 4) 
\item ``The scattering phases of individual partial waves, at a single
energy, can be determined by measurements of photoelectron angular
distributions. These are usually termed ``complete'' photoionization
experiments, and have been successfully demonstrated for a range of
atomic and molecular ionization process (see refs. \cite{Duncanson1976,Reid1992}
for example, for more comprehensive reviews see refs. \cite{Reid2003,Reid2012}),
and most recently for multi-photon ionization with femto-second pulses,
including electronic dynamics \cite{Hockett2014}.'' (Sect. 5) 
\item ``The reconstructed phases agreed reasonably well with theoretical
results, which were based on ePolyScat calculations similar to those
employed herein.'' (Sect. 5, in discussion of the results of ref.
\cite{Bertrand2013}, with ePolyScat calculations and post-processing
performed by some of the current authors) 
\item ``...calculations based on a modified 3-step model using time-dependent
ionization and propagation calculations, combined with accurate recombination
matrix elements (hence scattering phases) were able to recreate the
intensity envelope of the harmonic spectrum and spectral phase differences
between opposites end of the molecule'' (Sect. 5, in discussion of
the results of ref. \cite{Frumker2012} - again, ePolyScat calculations
and post-processing performed by some of the current authors) 
\item ``Molecular ionization is a complex phenomenon, with the outgoing
photoelectron wavepacket experiencing a highly anisotropic scattering
potential. In the time-domain, this results in a highly-structured
Wigner delay, as a function of energy and angle in the molecular frame.
With the use of scattering calculations, the angle-dependent Wigner
delay $\tau_{w}^{g}(k,\,\theta,\,\phi)$ was examined for two simple
diatomics, and these results illustrate the magnitudes of the delays,
and types of structures, which might generally be expected in molecular
photoionization. The deep link between the Wigner delay and the photoionization
matrix elements is also revealed in the correlation of energy-domain
photoionization phenomena - in this case the shape resonance in $N_{2}$
- with features in the Wigner delay.'' (Sect. 6)
\end{itemize}
\end{itemize}

\section*{Summary}

In conclusion, we hope our response to Baykusheva \& Wörner's contention
is quite clear and robust: there is no issue here, either conceptually,
in the technical details of the formalism used, or numerically. We
further hope that our lengthy and detailed response serves as useful
material for  researchers interested in the fascinating and challenging
topics of molecular photoionization and Wigner delays, and provides
a suitable reading list of canonical works.

\bigskip{}

\section*{Appendix - Notes on Scattering Theory \& ePolyScat for Photoionization,
reproduced from the Supp. Mat.}

The following notes, originally included in the Supplementary Materials
for our article (available online via Figshare at \url{http://dx.doi.org/10.6084/m9.figshare.2007486}),
are reproduced here for the interested reader.

\section*{General framework}

Within scattering theory the free particle wavefunction can be expressed
as a partial wave expansion in radial and angular functions:

\begin{equation}
\Psi(\mathbf{r})=\sum_{lm}A_{l}\chi_{l}(r)Y_{lm}(\theta,\phi)=\sum_{lm}\psi_{lm}(\mathbf{r})\label{eq:psi_scat}
\end{equation}

The exact form of $\chi_{l}(r)$ and $A_{l}$ will depend on the potential
$V(r)$. For a Coulombic potential, $V(r)\propto Z_{1}Z_{2}/r$, where
$Z_{1}$ and $Z_{2}$ are the charges on the scattering centre and
scattered particle, solutions are given by (using incoming wave normalization):

\begin{equation}
\chi_{l}(r)=F_{l}(r)\overset{{\scriptstyle r\rightarrow\infty}}{\longrightarrow}\sin\left[kr-\frac{\pi l}{2}-\frac{Z_{1}Z_{2}}{k}\ln(2kr)+\sigma_{l}\right]\label{eq:Fl_asymptote}
\end{equation}

\begin{equation}
A_{l}=\frac{2l+1}{kr}i^{l}e^{-i\sigma_{l}}
\end{equation}

\begin{equation}
\sigma_{l}=\arg\Gamma\left[l+1-i\frac{Z_{1}Z_{2}}{k}\right]\label{eq:coulomb-phase}
\end{equation}

Here the solution to the radial wavefunction, $\chi_{l}(r)$, is given
by the (regular) \emph{Coulomb function} $F_{l}(r)$. This has a complicated
functional form near the scattering centre, but asymptotically goes
to a sinusoidal form. Both $F_{l}$ and $A_{l}$ contain terms involving
$\sigma_{l}$, this is the \emph{Coulomb phase}, and is given by equation
\ref{eq:coulomb-phase}. $\Gamma$ is the gamma function.

While the Coulomb potential is the exact form for a point charge,
more generally a scattering system may have an additional short-range
contribution to the potential (one which scales as $1/r^{n}$, where
$n>1$), and this contribution may be non-centrally symmetric (i.e.
anisotropic). However, the strength of these short-range interactions
and multi-polar contributions to the potential will fall to zero much
faster than the Coulombic term, and we can define a boundary, $r_{c}$,
beyond which the potential is purely Coulombic. We now have a potential
defined by $V'(r<r_{c})$ and $V(r\geq r_{c})$. We do not know the
exact (analytic) form of the wavefunction in the region $r<r_{c}$.
However, in the Coulombic region the radial wavefunction still has
an analytic form, and is now described by:

\begin{equation}
\chi_{l}(r\geq r_{c})=\cos(\delta_{lm})F_{l}(r)+\sin(\delta_{lm})G_{l}(r)
\end{equation}

\begin{equation}
G_{l}(r)\overset{{\scriptstyle r\rightarrow\infty}}{\longrightarrow}\cos\left[kr-\frac{\pi l}{2}-\frac{Z_{1}Z_{2}}{k}\ln(2kr)+\sigma_{l}\right]
\end{equation}

\begin{equation}
A_{l}=\frac{2l+1}{kr}i^{l}e^{-i(\sigma_{l}+\delta_{lm})}
\end{equation}

Here $F_{l}(r)$ is the regular Coulomb function as before, while
$G_{l}(r)$ is the \emph{irregular Coulomb function}. $\delta_{lm}$
is an additional \emph{scattering phase shift}, which arises from
the non-Coulombic part of the scattering potential. This phase shift
defines the mixing of the regular and irregular Coulomb functions,
and this mixing also determines the asymptotic phase shift:

\begin{equation}
\chi_{l}\overset{{\scriptstyle r\rightarrow\infty}}{\longrightarrow}\sin\left[kr-\frac{\pi l}{2}-\frac{Z_{1}Z_{2}}{k}\ln(2kr)+\sigma_{l}+\delta_{lm}\right]
\end{equation}

Hence the scattering phase $\delta_{lm}$ describes the effect of
the non-Coulombic part of the potential, $V'(r)$, and is labelled
with $m$ to show that this may affect different components of each
$l$-wave differently in an anisotropic scattering system. Note that
the short-range part of the potential may still be centrally-symmetric,
in which case only $m=0$ components will be present. Although the
form of the wavefunction is not generally known for $r<r_{c}$ (but
could be found numerically for a given $V'(r)$), the scattering phase
carries all of the information on the stength of the short range potential.

Most generally, the overall phase of each partial-wave channel is
denoted $\eta_{lm}=\sigma_{l}+\delta_{lm}$. The total phase (including
angle-dependence) can be most cleanly written as, simply, $\eta_{t}(\mathbf{r})=arg(\Psi(\mathbf{r}))$,
which incorporates the scattering phases $\eta_{lm}$, as well as
any additional channel-dependent phase contributions (e.g. phase contributions
from $Y_{lm}$ terms etc.). 

\section*{Photoionization}

The asymptotic wavefunction defines the final state of the system,
thus any experimental observations. The solution above defines the
continuum wavefunction in the presence of the scattering potential.
In the case of photoionization, the amplitudes of the various partial-waves
in the asymptotic limit must thus be found from some overlap from
the initial state. Typically we work within the dipole regime, and
the light-matter coupling at an energy $E$ can be written:

\begin{equation}
d(E)=\langle\Phi_{f}(\mathbf{r});\:\Psi(\mathbf{r})|\hat{\mathbf{\mu}}.\boldsymbol{\mathbf{E}}|\Phi_{i}(\mathbf{r})\rangle\label{eq:dipole-int}
\end{equation}
Here $\Psi(\mathbf{r})$ is the continuum (photoelectron) wavefunction
of eqn. \ref{eq:psi_scat}; $\Phi_{i}(\mathbf{r})$ is the inital
$N$-electron state and $\Phi_{f}(\mathbf{r})$ the $N-1$ electron
final state of the ionizing molecule; $\hat{\mu}$ is the dipole operator
and $\mathbf{E}$ the incident electric field.

In terms of the observable asymptotic wavefunction (or photoelectron
wavepacket), which is defined as a function of energy and angle, these
matrix elements will determine the overall amplitudes and phases of
the continuum wavefunction prepared via photoabsorption. Hence we
can write the final asymptotic wavefunction/wavepacket as:

\begin{equation}
\Psi(E,\theta,\phi)=\sum_{lm}d_{l,m}(E)Y_{lm}(\theta,\phi)
\end{equation}
where $d_{l,m}(E)$ is the dipole matrix element expanded in partial-waves.
Here, the dipole matrix elements include radial integration over the
continuum wavefunction of eqn. \ref{eq:psi_scat}, and incorporates
the asymptotic phases $\eta_{lm}$. The final observable is the square
of this wavefunction, and (for an angle-sensitive measurement) will
retain phase sensitivity over the partial-wave channels:

\begin{equation}
I(E,\theta,\phi)=\Psi(E,\theta,\phi).\Psi^{*}(E,\theta,\phi)
\end{equation}

\section*{ePolyScat}

The dipole matrix elements are defined in ePolyScat \cite{Gianturco1994,Natalense1999,Lucchese2015}
by, e.g., the definition of the MF-PADs as per eqns. 1-3 of ref. \cite{Toffoli2007}:

\begin{equation}
I_{\mu_{0}}(\theta_{\hat{k}},\phi_{\hat{k}},\theta_{\hat{n}},\phi_{\hat{n}})=\frac{4\pi^{2}E}{cg_{p_{i}}}\sum_{\mu_{i},\mu_{f}}|T_{\mu_{0}}^{p_{i}\mu_{i},p_{f}\mu_{f}}(\theta_{\hat{k}},\phi_{\hat{k}},\theta_{\hat{n}},\phi_{\hat{n}})|^{2}\label{eq:MFPAD}
\end{equation}

\begin{equation}
T_{\mu_{0}}^{p_{i}\mu_{i},p_{f}\mu_{f}}(\theta_{\hat{k}},\phi_{\hat{k}},\theta_{\hat{n}},\phi_{\hat{n}})=\sum_{l,m,\mu}I_{l,m,\mu}^{p_{i}\mu_{i},p_{f}\mu_{f}}(E)Y_{lm}^{*}(\theta_{\hat{k}},\phi_{\hat{k}})D_{\mu,-\mu_{0}}^{1}(R_{\hat{n}})\label{eq:TMF}
\end{equation}

\begin{equation}
I_{l,m,\mu}^{p_{i}\mu_{i},p_{f}\mu_{f}}(E)=\langle\Psi_{i}^{p_{i},\mu_{i}}|\hat{d_{\mu}}|\Psi_{f}^{p_{f},\mu_{f}}\varphi_{klm}^{(-)}\rangle\label{eq:I}
\end{equation}

In this formalism:
\begin{itemize}
\item $I_{l,m,\mu}^{p_{i}\mu_{i},p_{f}\mu_{f}}(E)$ is the radial part of
the dipole matrix element, determined from the initial and final state
electronic wavefunctions $\Psi_{i}^{p_{i},\mu_{i}}$and $\Psi_{f}^{p_{f},\mu_{f}}$,
photoelectron wavefunction $\varphi_{klm}^{(-)}$ and dipole operator
$\hat{d_{\mu}}$. Here the wavefunctions are indexed by irreducible
representation (i.e. symmetry) by the labels $p_{i}$ and $p_{f}$,
with components $\mu_{i}$ and $\mu_{f}$ respectively; $l,m$ are
angular momentum components, $\mu$ is the projection of the polarization
into the MF (from a value $\mu_{0}$ in the LF). Each energy and irreducible
representation corresponds to a calculation in ePolyScat.
\item $T_{\mu_{0}}^{p_{i}\mu_{i},p_{f}\mu_{f}}(\theta_{\hat{k}},\phi_{\hat{k}},\theta_{\hat{n}},\phi_{\hat{n}})$
is the full matrix element (expanded in polar coordinates) in the
MF, where $\hat{k}$ denotes the direction of the photoelectron $\mathbf{k}$-vector,
and $\hat{n}$ the direction of the polarization vector $\mathbf{n}$
of the ionizing light. Note that the summation over components $\{l,m,\mu\}$
is coherent, and hence phase sensitive.
\item $Y_{lm}^{*}(\theta_{\hat{k}},\phi_{\hat{k}})$ is a spherical harmonic.
\item $D_{\mu,-\mu_{0}}^{1}(R_{\hat{n}})$ is a Wigner rotation matrix element,
with a set of Euler angles $R_{\hat{n}}=(\phi_{\hat{n}},\theta_{\hat{n}},\chi_{\hat{n}})$,
which rotates/projects the polarization into the MF .
\item $I_{\mu_{0}}(\theta_{\hat{k}},\phi_{\hat{k}},\theta_{\hat{n}},\phi_{\hat{n}})$
is the final (observable) MFPAD, for a polarization $\mu_{0}$ and
summed over all symmetry components of the initial and final states,
$\mu_{i}$ and $\mu_{f}$. Note that this sum can be expressed as
an incoherent summation, since these components are (by definition)
orthogonal.
\item $g_{p_{i}}$ is the degeneracy of the state $p_{i}$.
\end{itemize}
The dipole matrix element of eqn. \ref{eq:I} - the radial part of
the dipole matrix element - effectively defines the final state amplitude
and phase. Hence, is equivalent to the general form of eqn. \ref{eq:dipole-int},
but here expanded in terms of symmetries of the light-matter system.

In practice, the inital $N$-electron and final $(N-1)$-electron
wavefunctions are defined by standard computational chemistry methods
(as implemented in Gaussian, Gamess, etc.). The scattering state is
solved numerically by ePS via a Schwinger variational procedure \cite{Lucchese1986},
and the radial dipole integrals solved based on this scattering state.
Numerically, an effective range for the interaction ($r_{max}$) is
defined by the spatial grid used in the calculation; other calculation
parameters may also affect the numerical results, see ref. \cite{Lucchese1986}.
Matrix elements $I_{l,m,\mu}^{p_{i}\mu_{i},p_{f}\mu_{f}}(E)$ are
output for further processing, e.g. for MF-PADs or calculation of
Wigner delays.

\bibliographystyle{unsrtnat}
\bibliography{ionization_time_delay_081016,hockett_refs_111016,PADs-ePolyScat,tannor}

\end{document}